\documentclass[twocolumn, aps,raggedbottom,nobalancelastpage,amssymb,floatfix]{revtex4-1} %
\usepackage{graphicx}
\usepackage{amsmath}
\usepackage{amssymb}
\usepackage{bm}
\usepackage{pgf}
\usepackage{color}
\usepackage{amsfonts}
\usepackage{hyperref}
\usepackage{siunitx}
\usepackage{tabularx}

\begin{document}

\title{ A flexible spiraling-metasurface as a versatile haptic interface}

\author{Osama R. Bilal}\email{osama.bilal@uconn.edu}
\affiliation{Division of Engineering and Applied Science, California Institute of Technology, Pasadena, California 91125, USA}
\affiliation{Department of Mechanical Engineering, University of Connecticut, Storrs, Connecticut 06269, USA}

\author{Vincenzo Costanza}
\affiliation{Division of Engineering and Applied Science, California Institute of Technology, Pasadena, California 91125, USA}

\author{Ali Israr}
\affiliation{Facebook Inc., Menlo Park, CA 94025, USA}

\author{Antonio Palermo}
\affiliation{Division of Engineering and Applied Science, California Institute of Technology, Pasadena, California 91125, USA}

\author{Paolo Celli}
\affiliation{Division of Engineering and Applied Science, California Institute of Technology, Pasadena, California 91125, USA}

\author{Frances Lau}
\affiliation{Facebook Inc., Menlo Park, CA 94025, USA}

\author{Chiara Daraio}\email{daraio@caltech.edu}
\affiliation{Division of Engineering and Applied Science, California Institute of Technology, Pasadena, California 91125, USA}

\keywords{Metasurfaces, Haptic interface, Virtual and Augmented Reality, Actuators}

\maketitle
	\textbf{Haptic feedback is the most significant sensory interface following visual cues. Developing thin, flexible surfaces that function as haptic interfaces is important for augmenting virtual reality, wearable devices, robotics and prostheses. For example, adding a haptic feedback interface to prosthesis could improve their acceptance among amputees. State of the art programmable interfaces  targeting the skin feel-of-touch through mechano-receptors are limited by inadequate sensory feedback, cumbersome mechanisms or narrow frequency of operation. Here, we present a flexible metasurface as a generic haptic interface capable of producing complex tactile patterns on the human skin at wide range of frequencies. The metasurface is composed of multiple ``pixels" that can locally amplify both input displacements and forces. Each of these pixels encodes various deformation patterns capable of producing different sensations on contact. The metasurface can transform a harmonic signal containing multiple frequencies into a complex preprogrammed tactile pattern. Our findings, corroborated by user studies conducted on human candidates, can open new avenues for wearable and robotic interfaces.}\\

Haptic interfaces enable us to receive a tactile response to varying shapes and textures that might not be possible using visual feedback \cite{johansson2009coding}. These interfaces can be either touchable or wearable. Touch-based interfaces, like touchscreens, rely on the high density of receptors at the finger tips in the human hand \cite{chortos2016pursuing, johansson2009coding} to deliver haptic feedback. They operate, in most cases,  on electrostatic forces to control the friction between a user's finger and a screen \cite{banter2010touch,bau2010teslatouch,radivojevic2012electrotactile, meyer2013fingertip,ilkhani2018creating} or on exciting ultrasonic flexural waves, traveling on an elastic, uniform screen \cite{pance2013method}. Wearable interfaces pose more challenges in design, compared to touch-based ones, largely because of the relatively fewer number of available receptors. In  particular when designed for scarce receptor regions such as the human arm, back or leg. Electrostimulation through attached electrodes to the skin is one approach to overcome the scarcity of mechanoreceptors for wearable interfaces \cite{withana2018tacttoo}. Such methodology faces the challenge of variability in skin and its impedance between different body parts and people \cite{cogan2008neural}, in addition to the difficulty in selecting appropriate combinations of voltages and currents to create desired responses without pain or electrically induced lesions \cite{yu2019skin}. 

A different avenue to haptic stimulation is the utilization of mechanical forces at low frequencies to induce tactile sensation at the skin-haptic interface. However, the spatial resolution achievable and the force output reachable by such technology correlates strongly with the number of actuators. The limitation stems mainly from the required trade-off between the size, number and power consumption of the available actuators to reach complex tactile patterns. Such a pattern can be achieved by a grid of connected actuators. However, providing power and control for such a grid is a challenge. To that extent, haptic actuators can be divided into two categories; bulky and powerful - such as acoustic coils- or thin and weak - such as piezoelectric actuators. Moreover, most actuators are limited to a single polarization excitation (e.g., out-of-plane or in-plane) or limited frequency bandwidth. Therefore, the need for flexible, thin, yet scalable amplification mechanisms is apparent \cite{chortos2016pursuing,saal2015biomimetic,wijk2015forearm,nghiem2015providing,anikeeva2015restoring,kapur2010spatially,yu2019skin,berger2018uncanny}.

In this paper, we propose the utilization of metasurfaces \cite{liu2000locally,ma2016acoustic,bilal2017reprogrammable,foehr2018spiral,bilal2019haptic} as a versatile haptic interface. Our elastic metasurface consists of a flexible layer of structured unit cells, designed to include resonant elements that can be excited by low-power transducers. The operational frequencies of the metasurface can span multiple orders of magnitude by design. The use of local resonances within the metasurface, as opposed to flexural waves on a uniform, rigid surface allows to locally amplify a mechanical signal. Such amplification can enhance both the force amplitude and the resolution of the vibrotactile sensation on contact with the skin. Each unit cell is fabricated with a sub-millimeter thickness and acts as an individual tactile “pixel”. The “pixels” are designed to vibrate at the target frequencies with various deformation patterns, resulting in a plethora of vibrotactile sensations at the interface with the skin. Each unit cell in the metasurface can be tuned to respond to different resonance frequencies. When the surface is excited with a single-frequency harmonic signal, only a single resonator, or a small group of resonators, will activate and amplify the signal. When a harmonic signal consisting of several frequencies is used for excitation, it can actuate different resonators either in parallel or in series, producing a complex tactile pattern. The metasurfaces can then be incorporated in a multilayer device (for example, including a layer of pixel-addressed piezoactuators) and can be embedded in an insulating mounting platform (e.g., bracelet or T-shirt), to create an actuated, flexible haptic interface.

\begin{figure}[b]
	\begin{center}
		\includegraphics[scale =1]{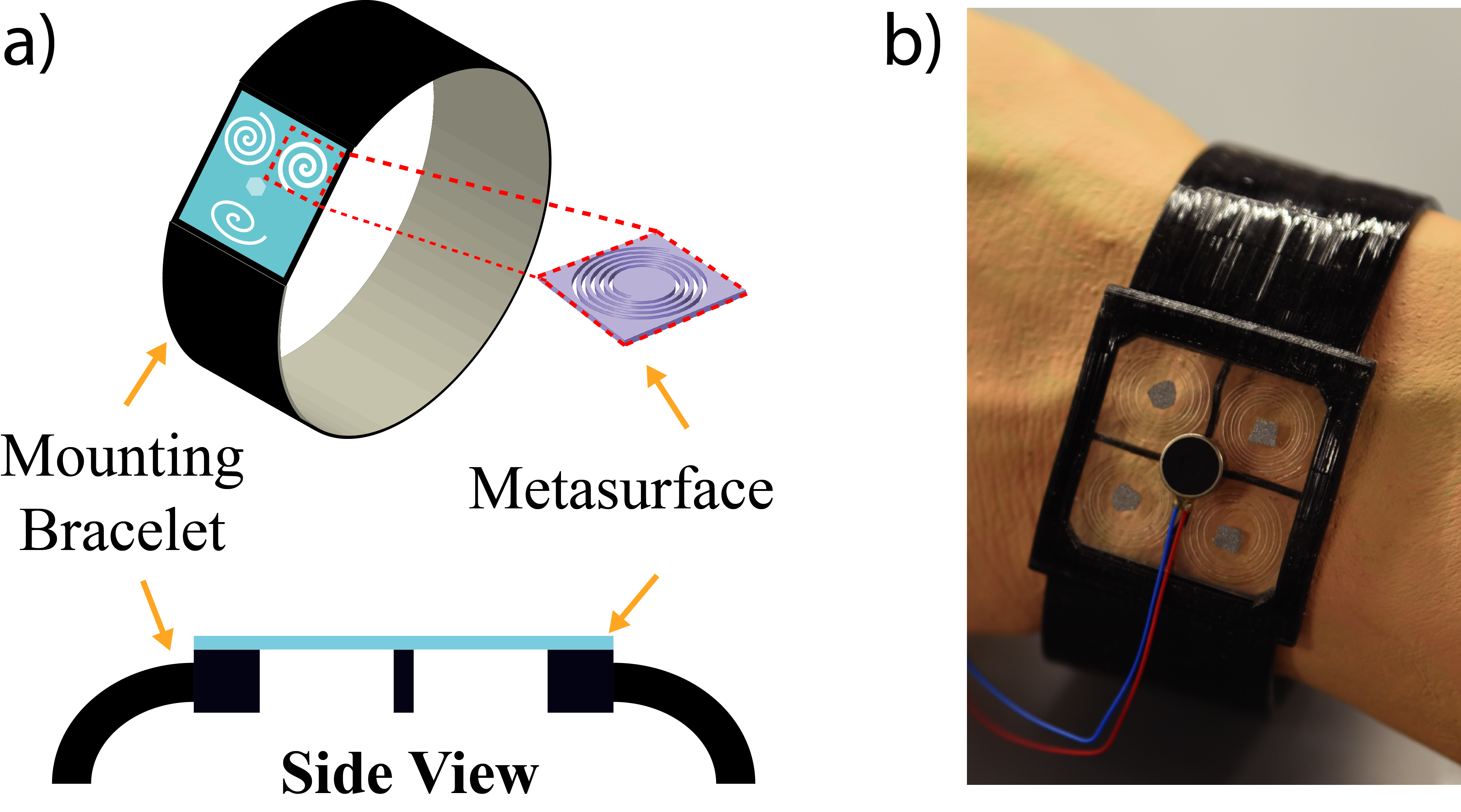}
	\end{center}
	\caption{\textbf{Metasurfaces as a haptic interface.} (a) A schematic of a metasurface mounted on an a bracelet. (b) A prototype of a metasurface with four pixels arranged in a 2x2 grid mounted on a 3D printed rubber bracelet. A linear resonant actuator (LRA) is attached at the center of the metasurface.}
	\label{fig:concept}
\end{figure}

\begin{figure*}[t]
	\begin{center}
		\includegraphics[scale =1]{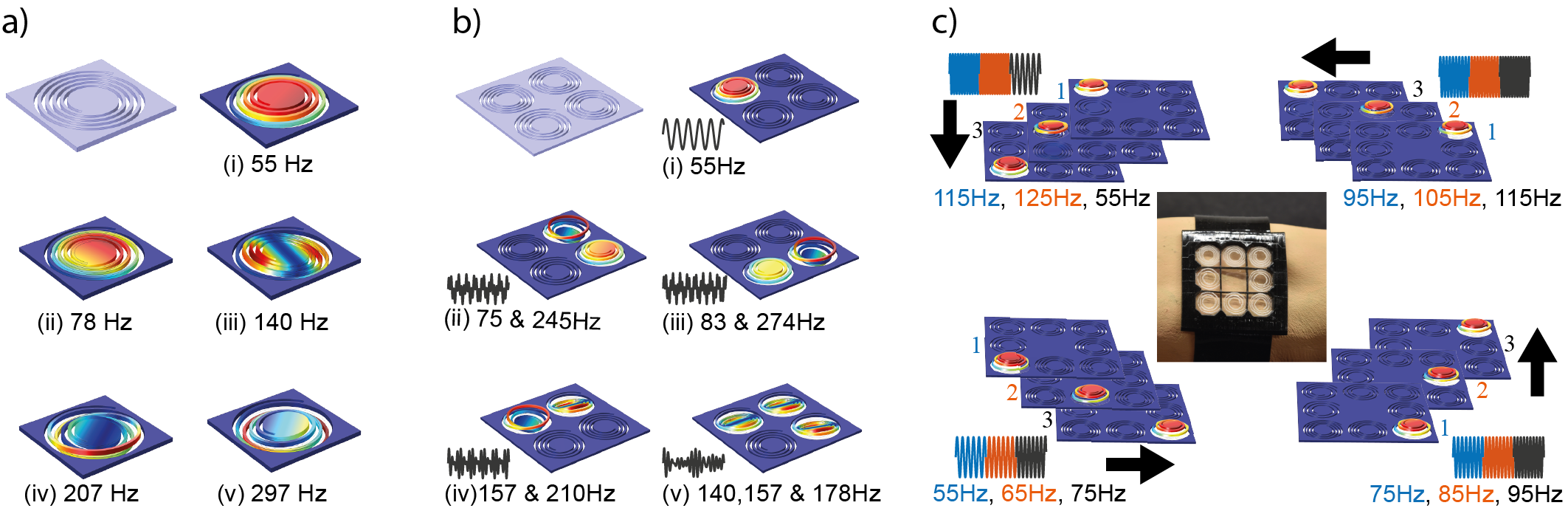}
	\end{center}
	\caption{\textbf{Metasurface pixels analysis}  (a) A single  pixel (top left) along with the numerical simulation of its first five mode shapes at different frequencies. (b) A grid of four pixels arranged in 2 $\times$ 2 square grid along with the numerical response of the metasurface when excited by one, two or three different frequencies that matches one or more of the mode shapes of the individual pixels.  (c) An eight-pixel metasurface mounted on a forearm with a rubber bracelet. The three mode shapes of the metasurface when excited by a signal containing the three resonant frequencies along one side to indicate the four principal directions.}
	\label{fig:pixels}
\end{figure*}

One of the simplest geometries to consider as a resonator is a cantilever beam. By elongating such a beam and coiling it up in a spiraling shape \cite{bilal2017bistable, bilal2017reprogrammable, foehr2018spiral, bilal2020enhancement}, one can achieve two important objectives. The first is to encode many different mode shapes within a narrow frequency range. The second is to utilize space efficiently. It is possible to realize cantilevered pixels with geometries other than a spiral (e.g., a web-like cell, a snow-flake cell, etc.). In the following examples, however, we will use the Archimedean spiral as the fundamental resonating unit. This allows targeting selected vibratory haptic receptors in the skin that operate within the range of 5-400 Hz \cite{abraira2013sensory, anikeeva2015restoring, chortos2016pursuing}, in a relatively compact space. With this approach, one can efficiently increase (i) the modal resolution, (ii) the frequency resolution and (iii) the spatial resolution of an actuator while amplifying its output force and displacement. The increase in modal resolution can be achieved by having a single pixel induce varying deformation patterns that translate to different vibrotactile sensations upon contact with the skin, while using a single polarization actuator. The increase in frequency resolution can be achieved by having a single pixel generate distinguishable tactile sensations at neighboring frequencies that are otherwise indistinguishable by the skin. The spatial resolution can be increased by having multiple pixels affect different locations on the skin while using a single actuator. Since these mode shapes take place at the natural frequencies of the pixel(s) (i.e., the resonating spirals), the amplitudes of both output forces and displacements are amplified. Figure \ref{fig:concept}a shows an example schematic of a  bracelet with an attached metasurface composed of three pixels. A physical prototype of a bracelet mounted on a human wrist with four pixels connected to a single actuator is presented in Fig. \ref{fig:concept}b.

To demonstrate the concept of the metasurface as a generic haptic interface, we first consider a pixel made out of a single Archimedean cut through a polycarbonate sheet with a thickness $th$ = 0.5 mm. A single pixel can consist of one (or multiple) Archimedean spiral cut. The pixel side length is $a$ = 13 mm.  The spiraling cut is represented mathematically in polar coordinates as $r(s)= R-(R-r)\,s$, $\phi (s) = 2\pi\, n\, s$, where $r$ is the inside radius, $R$ is the outside radius, $n$ is the number of turns and $s \in [0; 1]$. The spiral cut has a width of 0.5 mm, r = 0.2a, R = 0.475a, n  = 4.25.  We model the dynamics of the unit cell using the finite element method in COMSOL$^{\textregistered}$ multi-physics. We identify the operational frequencies of the pixel by calculating its resonant mode shapes. Each frequency has its unique deformation pattern that can translate to a different sensation upon contact with the skin (Figure \ref{fig:pixels}a). For example, the first mode shape, taking place at 55 Hz, is a vertical motion for the core of the pixel (Figure \ref{fig:pixels}a(i)), while the second mode shape at 78 Hz is a transverse mode. Both mode shapes take place at neighboring frequencies (Figure \ref{fig:pixels}a(ii)), however with vastly different deformation patterns (in-plane and out-of-plane). The third mode is a tilting motion around a horizontal center line passing through the pixel (Figure \ref{fig:pixels}a(iii)). The fourth and fifth modes are in-phase and out-of-phase vertical motions for the outer edge of the spiral (Figure \ref{fig:pixels}a(iv-v)). The design of the pixel (its size, deformation patterns and operational frequency range) can be easily tuned \cite{foehr2018spiral} to accommodate the contact point(s) with the human skin and its mechanoreceptors density.

By systematically altering the pixel geometry, one can sculpt the frequency and deformation patterns of the metasurface. A group of slightly modified pixels, tessellated in an arbitrary formation, can work collectively to generate a plethora of tactile patterns. For example, by varying the length of the spiral cut $n$ between 2.5 and 4.1, we can engineer deformation patterns (i.e., mode shapes) similar to the ones in (Figure \ref{fig:pixels}a) at shifted frequencies. We arrange the altered pixels in a 2 $\times$ 2 square grid (Figure \ref{fig:concept}b). The metasurface grid can be actuated with either one or multiple harmonic frequencies, using the same actuator, based on the desired tactile pattern. For example, by exciting the metasurface with a harmonic wave at a frequency $f$ = 55 Hz, only the top left pixel will oscillate at its natural frequency with a vertical motion of the pixel core (Figure \ref{fig:pixels}b(i)). Actuating a single pixel within the grid with a certain mode shape can follow the same principle. When a more complex tactile pattern is required, more than one pixel can be excited with a single mechanical signal containing multiple harmonic frequencies. For example, both the top-right and top-left pixels can be excited simultaneously. A signal containing 157 Hz + 210 Hz results in the activation of the outer ring of the top-left pixel and a tilting motion around the center of the top-right one (Figure \ref{fig:pixels}b (iv)). A similar tactile pattern engaging the right side of the metasurface can be induced with a signal consisting of 75 Hz + 245 Hz (Figure\ref{fig:pixels}b(ii)). Engaging the same mode shape at selected locations simultaneously within the metasurface is also possible (Figure\ref{fig:pixels}b(v)).


In addition to actuating the metasurface pixels individually or collectively using space and frequency as design dimensions, one can consider time as an extra design dimension for our metasurface. For example, we can activate a series of  different pixels in meaningful sequences to send certain information through the skin (e.g., navigation). To achieve such an objective, we design a metasurface with eight pixels arranged in a 3 $\times$ 3 grid with total dimensions of 42 mm $\times$ 42 mm $\times$ 0.5 mm (Figure \ref{fig:pixels}c). The pixels are designed to have the same first mode shape (vertical motion at the center of the pixel) with a 10 Hz separation in their resonant frequencies. The metasurface is mounted on a 3d-printed rubber bracelet. The combination of the bracelet and the metasurface can provide directional information to a user when excited with a time signal activating different spiral resonators in sequence. This feature could be used in vision-less navigations systems. For example, to communicate a left turn, an excitation signal of 1 second at 55 Hz, followed by another at 65 Hz and finally 75 Hz can be felt on the skin as a continuous stroke towards the left (Figure \ref{fig:pixels}c).
\begin{figure}
	\begin{center}
		\includegraphics[scale =1]{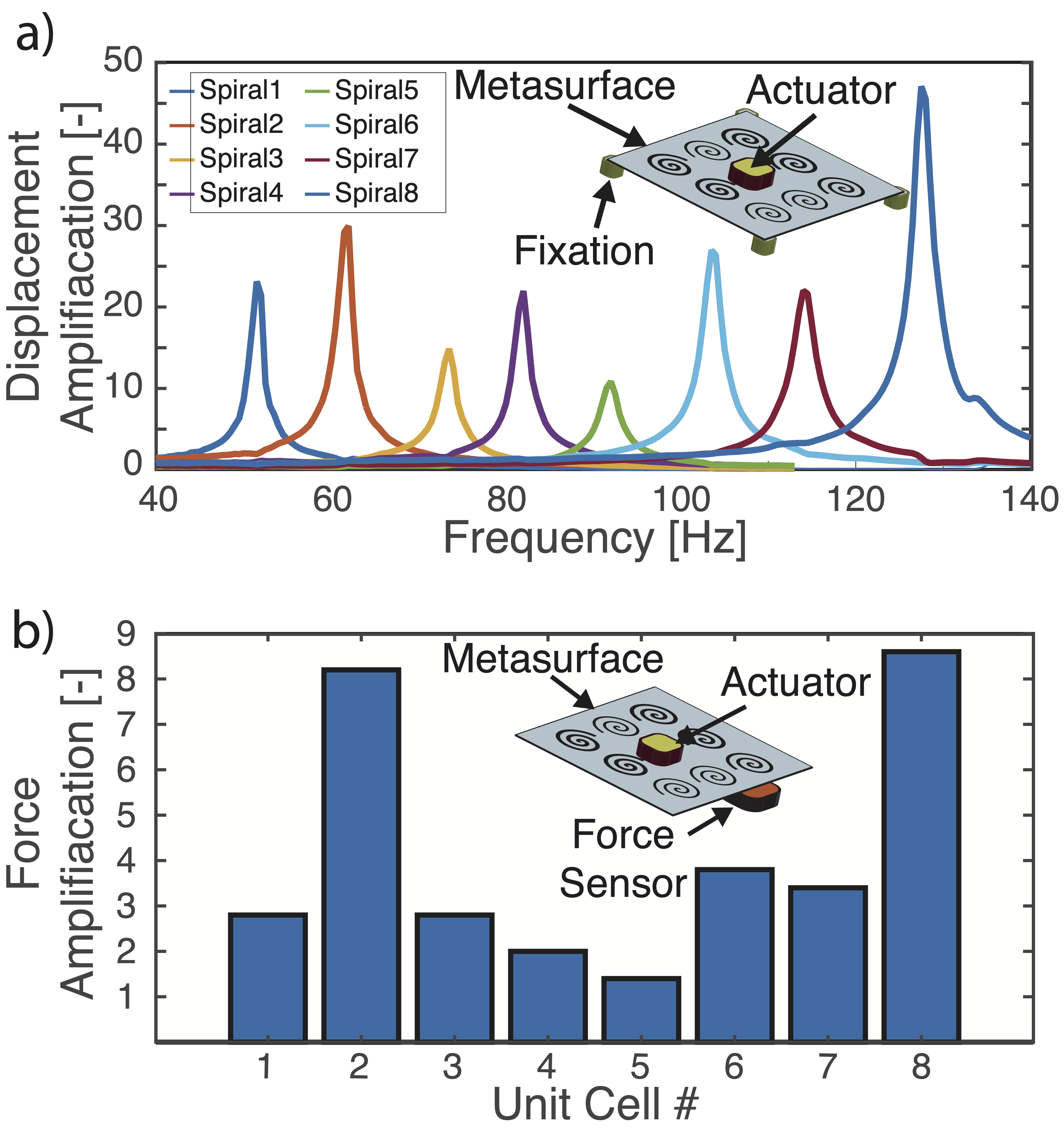}
	\end{center}
	\caption{\textbf{Input signal amplification:} (a) The displacement amplification (DA) factors as function of frequency in a square grid of 3$\times$3 pixels. DA is defined as the displacement measured at the center of each pixel normalized by the displacement measured at the attachment point of the actuator. (b) The force amplification (FA) factors as function of frequency in a square grid of 3$\times$3 pixels. FA is defined as the force measured at the center of each pixel normalized by the force measured at the attachment point of the actuator. The insets are schematic representations of the experimental setup to measure both displacement and force amplification.}
	\label{fig:amplification}
\end{figure}

\begin{figure}[b]
	\begin{center}
		\includegraphics[scale =1]{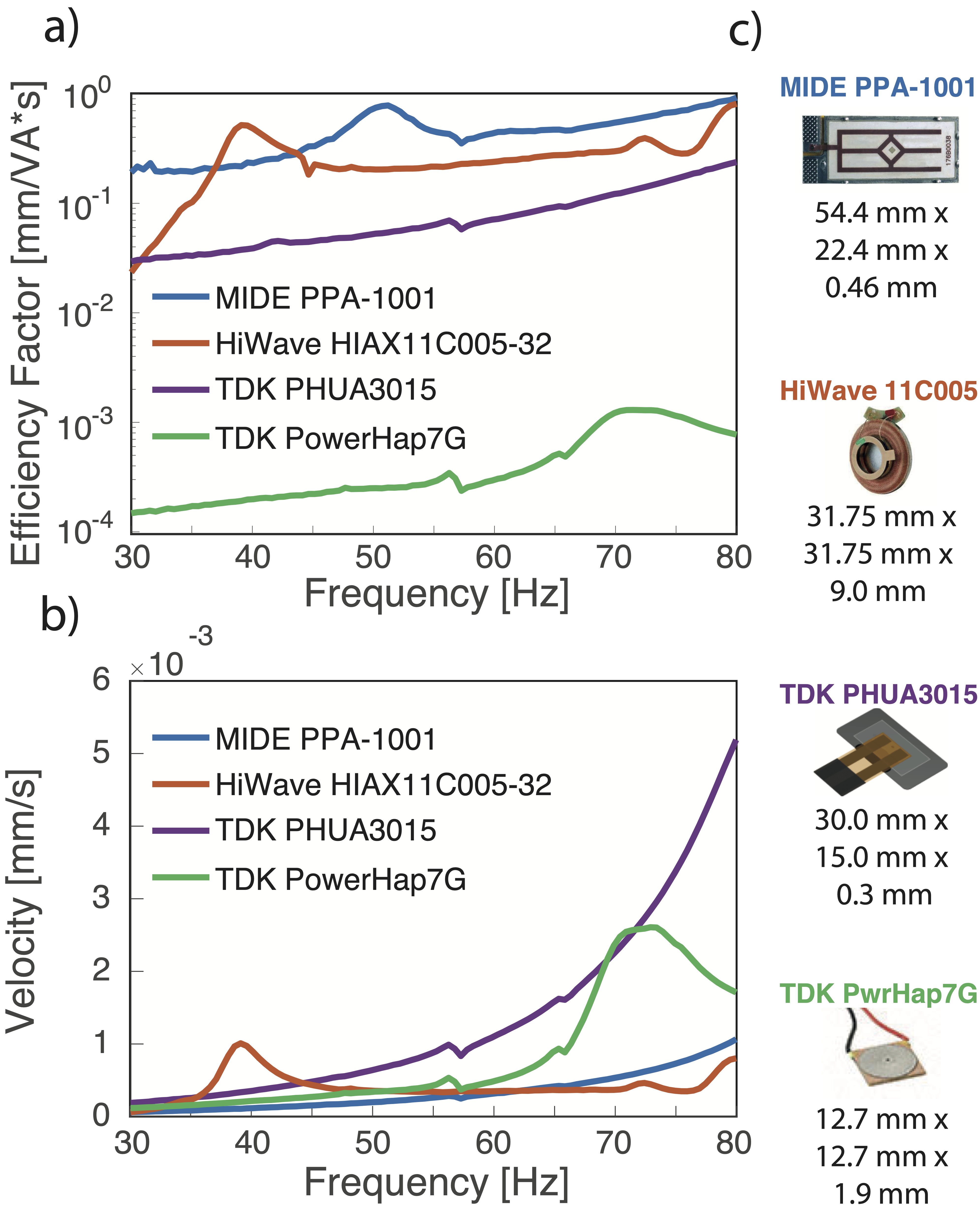}
	\end{center}
	\caption{\textbf{Actuators performance:} (a) Normalized efficiency factor of the different actuators calculated as velocity output relative to voltage input. (b) Measured velocity at the attachment point of each actuator to the metasurface. (c) A visual image of each actuator along with its dimensions. }
	\label{fig:actuators}
\end{figure}

To experimentally validate our numerical simulations, we first fabricate a 3 $\times$ 3 metasurface grid (Figure \ref{fig:pixels}c) with eight pixels (i.e., one pixel is missing at the center). We fix the four corners of the metasurface to cylindrical metal posts using double sided tape and attach an acoustic coil actuator (Hiax) at the center of the metasurface (Figure \ref{fig:amplification}a). We excite the metasurface with a harmonic, mechanical wave sweeping frequencies between 40 and 140 Hz. We record the out-of-plane displacement of the center of each pixel. We normalize the displacement of each pixel by the displacement amplitude at the center of the grid, where the actuator is attached, to calculate the amplification of displacement at each pixel site (Figure \ref{fig:amplification}a). The measurements show approximately a 10 Hz separation between the first mode of each of the eight pixels (i.e., the up and down motion of the pixel core) as calculated numerically. Moreover, as predicted, the pixels amplified the displacement amplitude by at least an order of magnitude at all sites (more than 40 times at some sites) (Figure \ref{fig:amplification}). The amplification is calculated as the ratio between the measured displacement at the actuator attachment point to the metasurface and the measured displacement at the pixel core.  It is important to note that larger displacement amplifications are easily realizable using our metasurfaces, however, such displacement amplitudes are too large to be measured using our laser Doppler vibrometer. In addition, we measure the force exerted by each pixel core upon contact with a piezoelectric force sensor (Figure \ref{fig:amplification}b). The harmonic force exerted by each pixel is measured using a calibrated Piezo disc at a fixed distance from the metasurface. In order to avoid overloading the sensor, we excite the metasurface using a low amplitude force. Then we measure the force signal at both the center of the metasurface (the attachment point of the actuator) and the center of each pixel. We normalize the measured force at each pixel center by the force measured at the center of the metasurface grid to calculate the amplification, in the same way we calculate the amplification of displacement. The metasurface also shows a clear amplification of the force amplitude compared to the force at the actuator attachment point. The amplification of both displacement and force takes place because resonances amplify displacement, velocity and acceleration. The variation in amplification amplitude stems from multiple factors, including the inherent resonance in the actuator as discussed later in the manuscript (Figure \ref{fig:actuators}), the position of the pixel relative to both actuator, boundary fixation and also the fabrication tolerances. However, despite variations, the experimental measurements  show clear amplification of the input mechanical signal. 

Actuation can be obtained with different systems, such as acoustic speakers, electromagnetic coils and piezoelectric transducers. Each actuator has its advantages and disadvantages. For example, acoustic coils work at very low voltage and can excite very high amplitude displacement at low frequencies, however, they are bulky. Piezoelectric transducers can be very thin and compact, however, they have a narrow operational bandwidth, require high voltage and most of the time break easily due to flexibility limitations. By combining our metasurfaces with some of these actuation mechanisms, one can envision overcoming their limitations. To demonstrate the validity of our proposal in conjunction with various types of actuation methods, we characterize the performance of our metasurfaces with four different actuators (Figure \ref{fig:actuators}). It should be noted that we are not comparing different actuators' performance against each other, but rather testing the efficiency and effectiveness of our metasurface with these different actuation methods. To perform the test, we fabricate a rectangular metasurface with five different pixels arranged in an ``L" shape configuration. The metasurface is fixed at its corners to four metal anchors. During the measurements, we attach the different actuators on top of the same metasurface, one-by-one in sequence, as shown in the inset of Figure \ref{fig:amplification}a. It is important to note that all the measurements are performed at very low amplitudes, however, larger amplification factors can be reached at higher excitation amplitudes. The tested actuators are (i) piezo actuator  Mide PPA-1001, (ii) acoustic coil HiWave hiax11c005-32, (iii) piezo actuator TDK PHUA3015 and (iv) piezo actuator TDK PowerHap7G. All actuators show clear mechanical signal amplification -above 27 folds- at 56 Hz (i.e., the first mode of spiral pixel). The amplification factors are 32, 34, 29 and 27 respectively, calculated as the ratio between the velocity at the center of the pixel divided by the velocity at each actuator attachment-point to the metasurface. We calculate the efficiency of the four tested actuators by normalizing their power consumption to the  measured velocities between 30 and 80 Hz (Figure \ref{fig:actuators} a). The calculated efficiency is an important factor to compare power consumption vs response. Such data could be used as a metric for choosing the appropriate actuator for a specific application. For example in a wearable device either Mide-PPA-1001 piezo or HiWave-11C005 acoustic coil gives the highest response for a given power input between 30 and 80 Hz. The acoustic coil shows a resonance peak at 39 Hz, with rapid decay in amplitude afterwards (Figure \ref{fig:actuators}b). The TDK-PHUA3015 shows a steadily increasing response throughout the plotted frequency range, but lower than the Mide piezo. Despite the very low response of all actuators at 56 Hz, the metasurface shows significant amplification. Such a result suggests the possibility of utilizing our metasurfaces at non-resonant frequencies of the actuators, which can increase their operational bandwidth dramatically. 


Our numerical analysis along with our experimental characterizations demonstrate the metasurface amplification of displacement, velocity, acceleration and force. Each individual pixel can be designed to have various mode shapes at targeted frequencies that deform differently. These pixels can be arranged in different configurations to produce arbitrary deformation patterns engaging one or more pixels at once or in sequence. However, the question remains open as to what degree such a concept can be effective on the human skin. To address this question, we perform user studies on human candidates. We start by considering a rather challenging spot on the human skin, the upper forearm, where skin is known to have a scarce amount of mechanoreceptors relative to the finger tips \cite{chortos2016pursuing, johansson2009coding}. To mount the metasurface on the skin, we fix it on a 3D printed rubber bracelet. The mounting bracelet secures a distance of at least 0.5 mm between the metasurface and the skin. This insulates the vibrations propagating through the entire metasurface and prevents them from being felt everywhere on the skin. To test the interface, we first carry out user study \# 1 where the metasurface is excited with either a resonant or a neighboring off-resonance frequency, at 62 Hz and 50 Hz, respectively. The users were asked to indicate which type of vibrations (or lack thereof) were felt on their forearm. The amplitude of both excitations is kept constant. The users were not allowed to look at the metasurface and wore noise canceling headphones to limit the feedback to the skin alone. The test included 120 trials on 4 different users. On average (Table \ref{table:usr1}), the users identified the correct type of excitation 94\% of the time (98\% when excited at resonant frequency and 90\% at non-resonant ones). This validates the haptic interface as a filter that only allows the preprogrammed resonant modes to make contact with the skin in a distinguishable manner.

\begin{table}[h]
	\caption{Summary of user study 1}
	\begin{tabular}{l|c|c}
		\hline
		Summary & Resonant freq. & Non-resonant freq. \\
		\hline 
		Resonant freq. & 98\% & 2\%   \\
		Non-resonant freq. & 10\% & 90\%
	\end{tabular}
	\label{table:usr1}
\end{table}



We carry out a second user study to test the discriminatory ability of the forearm skin to distinguish between neighboring pixels (separated by $\approx$ 14 mm) excited at neighboring frequencies (separated by 7 to 27 Hz). We use a metasurface with four pixels arranged in a 2 $\times$ 2 grid, similar to the one in Figure \ref{fig:concept}b. The distance between the center of each pixel is chosen to be 14 mm to accommodate the sparsity of the mechanoreceptors in the forearm \cite{chortos2016pursuing, johansson2009coding}. The frequencies are chosen to be very close to each other to test the limitation of our haptic interface. The first resonance mode of the each pixel is $f$ = \SI{52}{\hertz}, \SI{62}{\hertz}, \SI{69}{\hertz} and  \SI{79}{\hertz}. The metasurface is excited with two pulses. Each of these two pulses corresponds to the first resonant modes of one of the pixels. The two pulses last for one second each, separated by a pause of one second. Each pair of pulses is separated by a pause of 7 seconds before the following pair of pulses until the end of the study.  All frequencies are excited at the same amplitude. The user decides if the two pulses felt the same or different on the skin. The test included 288 trails on 6 different users. The users were able to discriminate between the two pulses (same vs. different) correctly 76\% of the time. The most difficult pulses to distinguish were between frequencies 2 and 3 (which are separated by only 7 Hz), with accuracy of 46\% (Table \ref{table:usr2}). Pulses that are separated by more than 17 Hz were the easiest to distinguish by more than 93\% accuracy, on average. The test results suggest the ability of the forearm skin to easily distinguish between two excitations separated by $\approx$\SI{14}{\mm} in space and \SI{17}{\hertz} in frequency, at an underutilized part of the human skin. If a more accurate response is required, one can increase the separation in either space or frequency, or both, by design.

\begin{table}
	\caption{Response summary for the different frequencies in user study 2}
	\begin{tabularx}{\columnwidth}{c|cccccc|c} \hline
		Freq. pair & ~~1-2~& ~~1-3 ~ & ~~1-4 ~& ~~2-3  ~& ~~2-4~& ~~3-4~& Confusion \\ 
		\hline 
		1-2 &  ~~75\%&  &  & & & & ~~~25\%\\
		1-3 &  &  ~92\%&  & & & & ~~~8\%\\
		1-4 &  &  &  ~96\% & & & & ~~~4\%\\
		2-3 &  &  &  & ~46\%& & & ~~~54\%\\
		2-4 &  &  &  & & ~92\% & & ~~~8\%\\
		3-4 &  &  &  & & & ~58\%& ~~~42\% 
	\end{tabularx}
	\begin{tabularx}{\columnwidth}{c|cccc|c}  \hline
		Freq.  pair& ~~~~1-1 ~~~ & ~~~~2-2 ~~~&~~~~3-3 ~~~& ~~~~4-4 ~~~& Confusion \\
		\hline
		1-1 & ~75\%&     	&        &     		& ~~~25\% \\
		2-2 &  	   & ~~75\% &      	 &     		& ~~~25\% \\
		3-3 &      &        & ~~~81\%&     		& ~~~19\% \\
		4-4 &      &        &     	 & ~~~75\%	& ~~~25\% 
	\end{tabularx}
	\label{table:usr2}
\end{table}

In this work, we present the design of a metasurface that can produce arbitrary deformation patterns by exciting selected resonant modes in suspended spiraling cantilevers. When placed in proximity to the skin, such platform can be used as a versatile haptic interface. The metasurface is composed of multiple ``pixels" that can locally amplify both input displacements and forces.  The versatility and scalability of the available tactile patterns  opens possibilities for a wide range of applications. For example, a metasurface grid composed of a few macro-pixels (patches of multiple pixels excited by a single actuator) can act as a wearable Braille communicator. Such platforms can also be used for virtual and augmented reality applications as well as prosthesis feedback. With enough resolution, one can induce realistic touch sensations on different parts of the human body.

\section*{Acknowledgment}
The authors are grateful for the fruitful discussions with Dr. Foehr and the financial support from Facebook Inc.

\section*{Materials and Methods}  The polycarbonate sheets have a measured thickness of \textit{th} = 0.5 mm, Young's modulus E = 2e9 Pa, density $\rho$ = 1200 Kg/m$^3$ and a bend radius $<$ 10 mm. Each pixel in the 2x2 and 3x3 configurations has a side length a = 13 mm and thickness th = 0.5 mm. The spiral cut has a width of 0.5 mm. The  inner radius of the spiral pattern is 2.6 mm and the outer radius is 6.175 mm. The specific number of turns for the spiraling pixels are \{ 2.6, 2.75, 2.91, 3.1, 3.3, 3.5, 3.8, 4.1\} turns. All the polycarbonate sheets are fabricated using standard CNC (computer numerical control) machining (model: PCV-60 50 TAPER). To characterize the metasurface in the user studies, we mount it in the TangoBlack bracelet. The bracelet is fabricated using a stratasys 500 printer. At the center of the metasurface we attach an acoustic coil HiWave-11C005 8 Ohms. The acoustic coil was the actuator of choice in the user study for its small footprint and low operational voltage. The acoustic coil is excited by a computer signal that passes through a Topping audio amplifier TP22. The out-of-plane displacement of the core of each pixel is measured using a Polytec laser Doppler vibrometer OFV-505 with a OFV-5000 decoder, using a VD-06 decoding card. The range of displacements that can be measured using the LDV is usually in the sub-millimeter scale. Therefore, we excite the metasurface with a small amplitude signal [10mV] at its center.

\bibliographystyle{unsrt}






\end{document}